\documentclass[preprint,authoryear]{elsarticle}

\usepackage[english]{babel}
\usepackage{natbib}
\usepackage[utf8]{inputenc}

\usepackage{amsmath}
\usepackage{amsfonts}

\usepackage{graphicx}
\graphicspath{{images/}}
\DeclareGraphicsExtensions{.pdf}

\usepackage{algpseudocode}
\usepackage{algorithm}

\usepackage{multirow,array}

\usepackage[usenames,dvipsnames]{xcolor}
\usepackage{color,soul}
\newcommand{\MOD}[1]{#1}

\usepackage{xspace}
\newcommand{\Ltwo}{\texttt{L2L2}\xspace}
\newcommand{\Lone}{\texttt{L2L1}\xspace}
\newcommand{\Lzero}{\texttt{L2L0}\xspace}
\newcommand{\DataDTIa}{$\mathtt{dti}_\mathtt{30}$\xspace}
\newcommand{\DataDTIb}{$\mathtt{dti}_\mathtt{20}$\xspace}
\newcommand{\DataHARDIa}{$\mathtt{hardi}_\mathtt{256}$\xspace}
\newcommand{\DataHARDIb}{$\mathtt{hardi}_\mathtt{50}$\xspace}
\newcommand{\DataHARDIc}{$\mathtt{hardi}_\mathtt{20}$\xspace}

\newcommand{\vect}[1]{\boldsymbol{#1}}

\usepackage[colorlinks=true]{hyperref}

\makeatletter
\def\ps@pprintTitle{%
  \let\@oddhead\@empty
  \let\@evenhead\@empty
  \def\@oddfoot{\reset@font\hfil\thepage\hfil}
  \let\@evenfoot\@oddfoot
}
\makeatother
\begin{document}

%%%%%%%%%%%%%%%%%%%%%%%%%
%%%%%  FrontMatter  %%%%%
%%%%%%%%%%%%%%%%%%%%%%%%% 
\begin{frontmatter}

\title{\MOD{Sparse regularization for fiber ODF reconstruction: from the suboptimality of $\ell_2$ and $\ell_1$ priors to $\ell_0$}}

\author[LTS5]{A. Daducci\corref{CORR}}
\ead{alessandro.daducci@epfl.ch}
%\ead[url]{http://people.epfl.ch/alessandro.daducci}
\author[MIPL,UNIGE]{D. Van De Ville}
\author[LTS5,UNIL]{J-P. Thiran}
\author[LTS5,MIPL,UNIGE,HW]{Y. Wiaux}

\address[LTS5]{Signal Processing Lab (LTS5), \'Ecole Polytechnique F\'ed\'erale de Lausanne, Switzerland}
\address[MIPL]{Medical Image Processing Lab, \'Ecole Polytechnique F\'ed\'erale de Lausanne, Switzerland}
\address[UNIL]{University Hospital Center (CHUV) and University of Lausanne (UNIL), Switzerland}
\address[UNIGE]{Department of Radiology and Medical Informatics, University of Geneva, Switzerland}
\address[HW]{Institute of Sensors, Signals \& Systems, Heriot-Watt University, Edinburgh, UK}
\cortext[CORR]{Corresponding author. \textit{Postal address:} EPFL STI IEL LTS5, ELD 232, Station 11, CH-1015 Lausanne (Switzerland). \textit{Phone:} +41 (0) 21 6934622}

\date{}

%%%%%%%%%%%%%%%%%%%%%%
%%%%%  Abstract  %%%%%
%%%%%%%%%%%%%%%%%%%%%% 
\begin{abstract}
Diffusion MRI is a well established imaging modality providing a powerful way to probe the structure of the white matter non-invasively. Despite its potential, the intrinsic long scan times of these sequences have hampered their use in clinical practice. For this reason, a large variety of methods have been recently proposed to shorten the acquisition times.
Among them, spherical deconvolution approaches have gained a lot of interest for their ability to reliably recover the intra-voxel fiber configuration with a relatively small number of data samples. To overcome the intrinsic instabilities of deconvolution, these methods use regularization schemes generally based on the assumption that the fiber orientation distribution (FOD) to be recovered \MOD{in each voxel} is sparse.
\MOD{The well known Constrained Spherical Deconvolution (CSD) approach resorts to Tikhonov regularization, based on an $\ell_2$-norm prior, which promotes a weak version of sparsity.
Also,} in the last few years compressed sensing has been advocated to further accelerate the acquisitions and $\ell_1$-norm minimization is generally employed as a means to promote sparsity in the recovered FODs.
In this paper, we provide evidence that the use of an $\ell_1$-norm \MOD{prior} to regularize this class of problems is somewhat inconsistent with \MOD{the fact that the fiber compartments all sum up to unity}.
To overcome this \MOD{$\ell_1$} inconsistency \MOD{while simultaneously exploiting sparsity more optimally than through an $\ell_2$ prior}, we reformulate the reconstruction problem as a constrained formulation between \MOD{a} data \MOD{term} and \MOD{and a sparsity prior consisting in an explicit bound on the $\ell_0$ norm of the FOD, i.e. on the number of fibers}.
The method has been tested both on synthetic and real data.
Experimental results show that \MOD{the proposed $\ell_0$ formulation} significantly reduces modeling errors \MOD{compared to the state-of-the-art $\ell_2$ and $\ell_1$ regularization approaches}.
\end{abstract}

%%%%%  Keywords  %%%%%

\end{frontmatter}

%%%%%%%%%%%%%%%%%%%%%%%%%%
%%%%%  Introduction  %%%%%
%%%%%%%%%%%%%%%%%%%%%%%%%% 
\section{Introduction} \label{Introduction}

Fiber-tracking is probably one of the most fascinating applications in diffusion MRI (dMRI), gathering a lot of attention since its introduction because of its ability to reconstruct the main neuronal bundles of the brain from the acquired data.
In fact, the random movement of the molecules in the white matter can be exploited for mapping brain connectivity, and structures otherwise invisible with other imaging modalities can be highlighted.
The study of this structural connectivity is of major importance in a fundamental neuroscience perspective, for developing our understanding of the brain, but also in a clinical perspective, with particular applications for the study of a wide range of neurological disorders.

The most powerful acquisition modality is diffusion spectrum imaging (DSI) \citep{Wedeen:2005aa}. It relies on cartesian signal sampling and is known to provide good imaging quality, but it is too time-consuming to be of real interest in a clinical perspective. Diffusion tensor imaging (DTI) \citep{Basser:1994aa} is always preferred instead. DTI is a very fast model-based technique providing valuable diagnostic information but, on the contrary, it is unable to model multiple fiber populations in a voxel. In a global connectivity analysis perspective, this constitutes a key limiting factor. Accelerated acquisitions relying on a smaller number of samples while providing accurate estimations of the intra-voxel fiber configuration thus represent an important challenge.

% spherical deconvolution
Recently, an increasing number of high angular resolution diffusion imaging (HARDI) approaches have been proposed for tackling this problem. In particular, spherical deconvolution (SD) based methods formed a very active area in this field \citep{Tournier:2004aa, Alexander:2005aa, Tournier:2007aa, Dellacqua:2007aa}. These methods rely on the assumption that the signal attenuation acquired with diffusion MRI can be expressed as the convolution of a given response function with the fiber orientation distribution (FOD). The FOD is a real-valued function on the unit sphere $\left(\mathrm{S}^2\right)$ giving the \textit{orientation} and the \textit{volume fraction} of the fiber populations present in a voxel. The response function, or kernel, describes the dMRI signal attenuation generated by an isolated single fiber population; it can be estimated from the data or represented by means of parametric functions.
SD approaches represented a big step in reducing the acquisition time of diffusion MRI, but are known to suffer heavily from noise and intrinsic instabilities in solving the deconvolution problem. For this reason, a regularization scheme is normally employed. A variety of approaches have been proposed, which are generally based on the two assumptions that the FOD is (i) a non-negative function and (ii) sparse, i.e. with only a few nonzero values, either explicitly or implicitly.
In fact, at the imaging resolution available nowadays, diffusion MRI is sensitive only to the major fiber bundles and it is commonly accepted that it can reliably \textit{disentangle up to 2-3 different fiber populations} inside a voxel \citep{Jeurissen:2010aa, Schultz:2012aa}. Hence, the FOD can reasonably be considered sparse in nature.
\MOD{In particular, the state-of-the-art Constrained Spherical Deconvolution (CSD) approach of \citet{Tournier:2007aa} resorts to Tikhonov regularization, based on an $\ell_2$-norm prior. While its primary purpose is to ensure the positivity of the FOD, it actually also implicitly promotes sparsity, but only a weak version of it.}

% compressed sensing
The recent advent of \emph{compressed sensing} (CS) theory \citep{Donoho:2006aa,Candes:2006aa,Baraniuk:2007aa} provided a mathematical framework for the reconstruction of sparse signals from under-sampled measurements mainly in the context of convex optimization.
CS has inspired new advanced approaches in the last few years for solving the reconstruction problem in diffusion MRI and allowed a further dramatic reduction in the number of samples needed to accurately infer the fiber structure in each voxel\MOD{, by promoting sparsity explicitly}.
For instance, \citet{Tristan-Vega:2011aa} and \citet{Michailovich:2011aa} recovered the orientation distribution function (ODF) by using different representations for the response function, while \citet{Merlet:2011aa} and \citet{Rathi:2011aa} focused on the full ensemble average propagator (EAP) of the diffusion process. In this work, however, we focus on spherical deconvolution based-methods and the quantity of interest is the FOD.
In general these methods are based on $\ell_1$ minimization, where the $\ell_1$ norm is defined as $|| \vect{x} ||_1 = \sum_i^n |\vect{x}_i|$ for any vector $\vect{x} \in \mathbb{R}^n$, and the common goal is to recover the FOD with fewest non-zeros that is compatible with the acquired dMRI data \citep{Ramirez-Manzanares:2007aa, Pu:2011aa, Landman:2012aa, Mani:2012aa}.
However, a minimum $\ell_1$-norm prior is inconsistent with the physical constraint that the sum of the volume fractions of the compartments inside a voxel is intrinsically equal to unity. %, i.e. $||x||_1 = 1$.

In this paper, we propose to exploit the versatility of compressed sensing and convex optimization to solve what we understand as \textit{the $\ell_1$ inconsistency}\MOD{, while simultaneously exploiting sparsity more optimally than the approaches based on the $\ell_2$ prior,} and improve the quality of FOD reconstruction in the white matter.
Our approach is as follows.
Strictly speaking, the FOD sparsity is the number of fiber populations, thus identified by the $\ell_0$ norm of the FOD.
$\ell_0$-norm problems are generally intractable as they are non convex, which explains the usual convex $\ell_1$-norm relaxation in the framework of compressed sensing. 
To this end, some greedy algorithms have been proposed to approximate the $\ell_0$ norm through a sequence of incremental approximations of the solution, such as Matching Pursuit~\citep{Mallat:1993aa} and Orthogonal Matching Pursuit~\citep{Pati:1993aa}.
However, the greedy and local nature of these algorithms, i.e. in the sense that compartments are identified sequentially, makes them suboptimal as compared to more robust approaches based on convex optimization, which are global in nature. In particular, a reweighted $\ell_1$ minimization scheme was developed by \citet{Candes:2008aa} in order to approach $\ell_0$ minimization by a sequence of convex weighted-$\ell_1$ problems.
We thus solve the $\ell_0$ minimization problem by making use of a reweighting scheme \MOD{and evaluate the effectiveness of the proposed formulation} in comparison with \MOD{state-of-the-art aprroaches based on either $\ell_2$ or $\ell_1$ priors}. We report results on both synthetic and real data.

%%%%%%%%%%%%%%%%%%%%%%%%%%%%%%%%%%%
%%%%%  Materials and methods  %%%%%
%%%%%%%%%%%%%%%%%%%%%%%%%%%%%%%%%%%
\section{Materials and methods} \label{MaterialsAndMethods}

%%%%%  Intra-voxel structure recovery via spherical deconvolution  %%%%%
\subsection{Intra-voxel structure recovery via spherical deconvolution} \label{Background}

As shown by \citet{Jian:2007aa}, spherical deconvolution methods can be cast into the following computational framework:
\begin{equation} \label{eqn:DeconvUnifiedFramework}
S(b,\vect{\hat{q}})/S_0 = \int R_{\vect{\hat{q}}}(\vect{\hat{p}}) \; f(\vect{\hat{p}}) \; d\Omega(\vect{\hat{p}}) ,
\end{equation}
where $f$ is the FOD to be estimated, $R_{\vect{\hat{q}}}$ the response function rotated in direction $\vect{\hat{q}} \in \mathrm{S}^2$ and the integration is performed over the unit sphere with $\vect{\hat{p}} = (\phi,\theta) \in \mathrm{S}^2$ and $d\Omega = \sin{\phi} \, d\phi \, d\theta$.
$S(b,\vect{\hat{q}})$ represents the dMRI signal measured on the q-space shell acquired with b-value $b$ in direction $\vect{\hat{q}} \in \mathrm{S}^2$, while $S_0$ is the signal acquired without diffusion weighting.
\MOD{The FOD $f$ is normally expressed as a linear combination of basis functions, e.g. spherical harmonics, as $f(\vect{\hat{p}}) = \sum_j w_j f_j(\vect{\hat{p}})$.}
The measurement process can thus be expressed in terms of the general formulation:
\begin{equation} \label{eqn:Recon-Problem}
\vect{y} = \Phi \vect{x} + \eta ,
\end{equation}
where $\vect{x} \in \mathbb{R}_+^n$ are the coefficients of the FOD, $\vect{y} \in \mathbb{R}^m$ is the vector with the dMRI signal measured in the voxel with $y_i = S(b,\vect{\hat{q}}_i)/S_0$ for $i \in \{ 1, \ldots, m \}$, $\eta$ represents the acquisition noise and \MOD{$\Phi = \{ \phi_{ij} \}\in \mathbb{R}^{m \times n}$ is the observation matrix modeling explicitly the convolution operator with the response function $R$, with $\phi_{ij} = \int R_{\vect{\hat{q}_i}}(\vect{\hat{p}}) \, f_j(\vect{\hat{p}}) \, d\Omega(\vect{\hat{p}})$.
Several choices for the convolution kernels and basis functions exist in the literature; more details will be provided on the specific $\Phi$ used with each algorithm considered in this work.}

\subsection{\MOD{\texorpdfstring{$\ell_2$}{Lg} prior}}
In the original formulation of \citet{Tournier:2004aa}, the FOD $\vect{x}$ and the measurements $\vect{y}$ were expressed by means of spherical harmonics (SH), and the deconvolution problem was solved by a simple matrix inversion. To reduce noise artifacts, a low-pass filter was applied for attenuating the high harmonic frequencies.
The method was improved in \citet{Tournier:2007aa} by reformulating the problem as an iterative procedure where, at each iteration, the current solution $\vect{x}^{(t)}$ is used to drive to zero the negative amplitudes of the FOD at the next iteration with a Tikhonov regularization \citep{Tikhonov:1977aa}:
\begin{equation} \label{eqn:SD-Problem}
\vect{x}^{(t+1)} = \underset{\vect{x}}{\operatorname{argmin}} \; || \Phi \vect{x} - \vect{y} ||_2^2 + \lambda^2 || L^{(t)} \vect{x}||_2^2 ,
\end{equation}
where $||\cdot||_p$ are the usual $\ell_p$ norms in $\mathbb{R}^n$, the free parameter $\lambda$ controls the degree of regularization and $L^{(t)}$ can be understood as a simple binary mask preserving only the directions of negative or small values of $x^{(t)}$. The $\ell_2$-norm regularization term therefore tends to send these values to zero, as probably spurious, hence \MOD{favoring large positive} values. Interestingly, %beyond sending negative values to zero
\MOD{beyond the claimed purpose of enforcing positivity}
, the operator $L$ thus \MOD{also} implicitly promotes a weak version of sparsity. However, this \MOD{$\ell_2$ prior} does not \MOD{explicitly} guarantee either positivity or sparsity in the recovered FOD.
In \citet{Alexander:2005aa} a maximum-entropy regularization was proposed to recover the FOD as the function that exhibits the minimum information content. The method showed higher robustness to noise than previous approaches, but was limited by the very high computational cost and did not promote sparsity.
Other regularization schemes have been proposed in the literature, but FOD sparsity has never been addressed with a rigorous mathematical formulation.

\subsection{\MOD{\texorpdfstring{$\ell_1$}{Lg} prior}}
Compressed sensing provides a powerful mathematical framework for the reconstruction of sparse signals from a low number of data \citep{Donoho:2006aa,Candes:2006aa}, mainly in the context of convex optimization.
According to this theory, it is possible to recover a signal from fewer samples than the number required by the Nyquist sampling theorem, provided that the signal is sparse in some \emph{sparsity basis} $\Psi$.
Let $\vect{x} \in \mathbb{R}^n$ be the signal to be recovered from the $m \ll n$ linear measurements $\vect{y} = \Phi \vect{x} \in \mathbb{R}^m$ and $\vect{\alpha} \in \mathbb{R}^n$ a sparse representation of $\vect{x}$ through $\Psi \in \mathbb{R}^{n \times n}$. If the observations $\vect{y}$ are corrupted by noise and $\Phi$ obeys some randomness and incoherence conditions, then the signal $\vect{x} = \Psi \vect{\alpha}$ can be recovered by solving the convex $\ell_1$ optimization problem:
\begin{equation} \label{eqn:CSProblem}
\underset{\vect{\alpha}}{\operatorname{argmin}} \; || \vect{\alpha} ||_1 \;\; \textrm{subject to} \;\; || \Phi \, \Psi \, \vect{\alpha} - \vect{y} ||_2 \le \epsilon ,
\end{equation}
where $\epsilon$ is a bound on the noise level.
\MOD{Assuming Gaussian noise, the square $\ell_2$ norm of the residual represents the log-likelihood of the data and follows a $\chi^2$ distribution. For a sufficiently large number of measurements, this distribution is extremely concentrated around its mean value. This fact is related to the well-known phenomenon of concentration of measure in statistics. Consequently, $\epsilon$ can be precisely defined by the mean of the $\chi^2$.}

In the context of FOD reconstructions, the sparsity basis $\Psi$ boils down to the identity matrix, thus $\vect{x} = \vect{\alpha}$.
In \citet{Ramirez-Manzanares:2007aa} and \citet{Jian:2007aa} the sensing basis $\Phi$, also called \emph{dictionary}, is generated by applying a set of rotations to a given Gaussian kernel (i.e. diffusion tensor) and the sparsest coefficients $\vect{x}$ of this linear combination best matching the measurements $\vect{y}$ are recovered by solving the following constrained minimization problem:
\begin{equation} \label{eqn:BPDNProblem}
\underset{\vect{x} \ge 0}{\operatorname{argmin}} \; || \vect{x} ||_1 \;\; \textrm{subject to} \;\; || \Phi \, \vect{x} - \vect{y} ||_2 \le \epsilon ,
\end{equation}
where the positivity constraint on the FOD values was directly embedded in the formulation of the convex problem.
\MOD{For $\mathrm{SNR}>2$ the noise in the magnitude dMRI images can be assumed Gaussian-distributed\footnote{Since dMRI data is commonly normalized by the baseline $S_0$, the $S_0$ image must be accurately estimated in order to keep the same noise statistics of the non-normalized signal. This is normally the case, though, as multiple $S_0$ volumes are commonly acquired in practice.} \citep{Gudbjartsson:1995aa}.
However, a statistical estimation of $\epsilon$ is not reliable, precisely because the number of measurements is very small and the $\chi^2$ is not really concentrated around its mean value. Thus, $\epsilon$ becomes an arbitrary parameter of the algorithm.
At very low SNR, one can also extrapolate the choice of the $\ell_2$ norm as a simple penalization term independent of statistical considerations.}

\MOD{The reconstruction problem can also be re-formulated as} a regularized (as opposed to constrained) $\ell_1$ minimization as in \citet{Landman:2012aa} and \citet{Pu:2011aa}:
\begin{equation} \label{eqn:1-normProblemRegularized}
\underset{\vect{x} \ge 0}{\operatorname{argmin}} \; || \Phi \vect{x} - \vect{y} ||_2^2 \; + \; \beta \; || \vect{x} ||_1 ,
\end{equation}
where the free parameter $\beta$ controls the trade-off between the data and the sparsity constraints.
In general, $\beta$ depends on the acquisition scheme and the noise level and it must be empirically optimized.
\MOD{Following the general CS approach, problems \eqref{eqn:BPDNProblem} and \eqref{eqn:1-normProblemRegularized} consider an $\ell_1$-norm prior on the FOD $\vect{x}$. However, in the dMRI context, a minimum $\ell_1$-norm prior is inconsistent with the physical constraint that the sum of the volume fractions of the compartments inside a voxel is intrinsically equal to unity, i.e. $|| \vect{x} ||_1 \equiv \sum_i \vect{x}_i = 1$.}
For this reason, we reckon that \MOD{also} these $\ell_1$-based formulations are intrinsically suboptimal.
\MOD{Fig.~\ref{fig:SumX} illustrates this inconsistency by reporting the $\ell_1$ norm of reconstructed FODs as a function of the amplitude of measurement noise.}

\MOD{Our main goal in this work is to demonstrate the suboptimalities of the approaches based on $\ell_2$ and $\ell_1$ priors and to suggest a new formulation, based on an $\ell_0$ prior, adequately characterizing the actual sparsity lying in the FOD.}

%%%%%  Reweighted $\ell_1$ minimization  %%%%%
\subsection{\MOD{\texorpdfstring{$\ell_0$}{Lg} prior}} \label{ReweightedL1Minimization}

In the aim of adequately characterizing the FOD sparsity, we re-formulate the reconstruction problem as a constrained $\ell_0$ minimization problem:
\begin{equation} \label{eqn:0-normProblem}
\underset{\vect{x} \ge 0}{\operatorname{argmin}} \; || \Phi \vect{x} - \vect{y} ||_2^2 \;\; \textrm{subject to} \;\; || \vect{x} ||_0 \leq k,
\end{equation}
where $||\cdot||_0$ explicitly counts the number of nonzero coefficients and $k$ represents \MOD{an upper bound on} the expected number of fiber populations in a voxel.

As already stated, the $\ell_0$ problems as such are intractable. The reweighting scheme proposed by \citet{Candes:2008aa} proceeds by sequentially solving weighted $\ell_1$ problems of the form \eqref{eqn:0-normProblem}, where the $\ell_0$ norm is substituted by a weighted $\ell_1$ norm defined as $||\vect{w} \vect{\alpha}||_1 = \sum_i \; \vect{w}_i \, |\vect{\alpha}_i|$, for positive weights $\vect{w}_i$ and where $i$ indexes vector components.
At each iteration, the weights are set as the inverse of the values of the solution of the previous problem, i.e. $\vect{w}_i^{(t)} \approx 1 / \vect{x}_i^{(t-1)}$.
At convergence, this set of weights makes the weighted $\ell_1$ norm independent of the precise value of the nonzero components, thus mimicking the $\ell_0$ norm while preserving the tractability of the problem with convex optimization tools.
Of course, it is not possible to have infinite weights for null coefficients; so a stability parameter $\tau$ must be added to the coefficients in the selection of the weights.

\begin{algorithm}[htbp]
\caption{Reweighted $\ell_1$ minimization for intra-voxel structure recovery}
\label{alg:ReweightedL1}
\small
\begin{algorithmic}
    \Require
    \hspace{0.5cm}Diffusion MRI signal $\vect{y} \in \mathbb{R}^m$ and sensing basis $\Phi \in \mathbb{R}^{m \times n}$
    \Ensure
    \hspace{0.2cm}FOD $\vect{x} \in \mathbb{R}^n$
    \State Set the initial status: \\
    	\hspace{0.5cm} $t \gets 0$ \hspace{0.1cm} and \hspace{0.1cm} $\vect{w}_i^{(0)} \gets 1, \;\; i = 1, \ldots, n$ \hspace{1cm} (the symbol $\gets$ denotes assignment)
    \Repeat
	    \State Solve the problem: \\
	    \hspace{1.0cm}$\vect{x}^{(t)} \gets \underset{\vect{x} \ge 0}{\operatorname{argmin}} \; || \Phi \vect{x} - \vect{y} ||_2^2 \;\; \textrm{subject to} \;\; || \vect{w}^{(t)} \vect{x} ||_1 \leq k$
	    \State Update the weights: \\
	    \hspace{1.5cm} $\vect{w}_i^{(t+1)} = \frac{1}{|\vect{x}_i^{(t)}| + \tau}$
	    \State $t \gets t + 1$
    \Until{stopping criterion is satisfied}
    \State $\vect{x} \gets \vect{x}^{(t-1)}$
\end{algorithmic}
\end{algorithm}

The \textit{main steps of the reweighted scheme} are reported in the algorithm \ref{alg:ReweightedL1}; in the remaining of the manuscript we will refer to it as \Lzero, as it is based on a $\ell_0$ prior.
We empirically set $\tau = 10^{-\MOD{3}}$ and the procedure was stopped if $\frac{||\vect{x}^{(t)} - \vect{x}^{(t-1)}||_1}{||\vect{x}^{(t-1)}||_1} < 10^{-3}$ between two successive iterations or after 20 iterations.
At the first iteration the weighted $\ell_1$ norm is the standard $\ell_1$ norm given $\vect{w}=1$, and therefore the constraint $|| \vect{w}^{(0)} \vect{x} ||_1 \leq k$ is a weak bound on the sum of the fiber compartments and does not constitute a limitation in the procedure.

\MOD{The proposed $\ell_0$ approach thus strongly promotes sparsity (by opposition with the $\ell_2$ approach) and circumvents the $\ell_1$ inconsistency. It is noteworthy that our formulation at least partially addresses the problem of arbitrary parameters such as $\epsilon$ in \eqref{eqn:BPDNProblem} and $\beta$ in \eqref{eqn:1-normProblemRegularized}, or $\lambda$ in \eqref{eqn:SD-Problem}. Our parameter $k$ indeed explicitly identifies an upper bound on the number of fibers.}
As discussed before and largely assumed in the literature, we can expect to have at maximum 2-3 fiber compartments in each voxel. The algorithm was found to be quite robust to the choice of $k$, and differences were not observed for values up to $k = 5$.

Finally, an explicit constraint $\sum_i \vect{x}_i = 1$ might have been added, as it represents the physical property that the volume fractions must sum up to unity.
For the sake of simplicity, in this work this constraint was not included (as it is always the case), assuming it is carried over by the data and well-designed bases as pointed out by \citet{Ramirez-Manzanares:2007aa}. \MOD{In sections \ref{VolumeFractionsAndL1} and \ref{RealDataVolumeFractionsAndL1}} we will provide evidence that actually this physical constraint is not met when using \MOD{$\ell_2$ or $\ell_1$ priors}, whereas it is correctly satisfied with our \MOD{proposed} $\ell_0$ formulation. This might have severe consequences on the reconstruction quality.

%%%%%  Comparison framework  %%%%%
\subsection{Comparison framework}

\MOD{We compared our $\ell_0$ approach based on problem \eqref{eqn:0-normProblem} against state-of-the-art $\ell_2$ and $\ell_1$ approaches respectively based on problems \eqref{eqn:SD-Problem} and \eqref{eqn:1-normProblemRegularized}, and referred to as \Ltwo and \Lone.}
To run \Ltwo reconstructions we made use of the original \texttt{mrtrix} implementation of \cite{Tournier:2012aa}, setting the optimal parameters as suggested by the software itself. To solve the \Lone and \Lzero problems we used the SPArse Modeling Software (SPAMS)\footnote{http://spams-devel.gforge.inria.fr}, an open-source toolbox written in C++ for solving various sparse recovery problems. SPAMS contains a very fast implementation of the LARS algorithm \citep{Efron:2004aa} for solving the LASSO problem and its variants as \MOD{the \Lone problem in equation \eqref{eqn:1-normProblemRegularized} and the weighted $\ell_1$ minimizations required for our \Lzero approach in equation \eqref{eqn:0-normProblem}}.
Numerical simulations on synthetic data were performed to quantitatively assess the performance of \Ltwo, \Lone and \Lzero under controlled conditions. \MOD{The effectiveness of the three priors was also assessed in case of real human brain data}.

%%%%%  Numerical simulations  %%%%%
\subsection{Numerical simulations} \label{SyntheticData}
Independent voxels with two fiber populations crossing at specific angles ($30^\circ - 90^\circ$ range) and with equal volume fractions were synthetically generated. The  signal $S$ corresponding to each voxel configuration was simulated by using the exact expression given in \citet{Soderman:1995aa} for the dMRI signal attenuation from particles diffusing in a restricted cylindrical geometry of radius $\rho$ and length $L$ with free diffusion coefficient $D_0$. The following parameters were used \citep{Ozarslan:2006aa,Jian:2007aa}: $L = 5  \; \mathrm{mm}$, $\rho = 5 \; \mathrm{\mu m}$, $D_0 = 2.02 \; \times \; 10^{-3} \; \mathrm{mm^2/s}$, $\Delta = 20.8 \; \mathrm{ms}$, $\delta = 2.4 \; \mathrm{ms}$.
The signal $S$ was contaminated with \emph{Rician noise} \citep{Gudbjartsson:1995aa} as follows:
\begin{equation}
S_{\mathrm{noisy}} = \sqrt{(S+\xi_1)^2 + (\xi_2)^2} ,
\end{equation}
where $\xi_1,\xi_2 \sim \mathcal{N}(0,\sigma^2)$ and $\sigma = S_0 / \mathrm{SNR}$ corresponds to a given signal-to-noise ratio on the $S_0$ image. We assumed $S_0 = 1$ without loss of generality.
Because of this assumption, we have implicitly considered a constant echo-time for acquisitions with different b-values, thus ignoring the fact that higher b-values normally require longer echo-times and therefore the images have a lower signal-to-noise ratio.
The study of the impact of the echo-time on different regularization priors is beyond the scope of our investigation.

For each voxel configuration, the signal was simulated at different b-values, $b \in \left\{ 500, 1000, \ldots, 4000 \right\} \mathrm{s/mm^2}$, and seven q-space sampling schemes were tested, respectively with 6, 10, 15, 20, 25, 30 and 50 samples equally distributed on half the unit sphere using electrostatic repulsion \citep{Jones:1999aa} assuming antipodal symmetry in diffusion signal.
Six different noise levels were considered, $\mathrm{SNR} = {5, 10, \ldots, 30}$. For every SNR, 100 repetitions of the same voxel were generated using different realizations of the noise. In our experiments, the actual signal-to-noise ratio in the simulated signal was always in a range where the Gaussian assumption on the noise holds. In the extreme setting with a $\textrm{SNR} = 5$ on the $S_0$ and $b = 4000 \; \mathrm{s/mm^2}$ the actual signal-to-noise ratio in the diffusion weighted signal was about 1.4.

%%%%%  Evaluation criteria  %%%%%
\subsection{Evaluation criteria} \label{EvaluationCriteria}
As one of the aims of this work is to improve SD reconstructions, we adopted standard metrics widely used in the literature \citep{Ramirez-Manzanares:2008aa,Landman:2012aa,Michailovich:2011aa} to assess the quality of the reconstructions with respect to number and orientation of the fiber populations:

\begin{itemize}

%  METRICs 1, 2 and 3
\item \emph{Probability of false fiber detection}. This metric quantifies the correct assessment of the real number $M$ of populations inside a voxel:
\begin{equation}
\mathrm{P_d} = \frac{ | M - \tilde{M} | }{ M } \cdot 100 \%,
\end{equation}
where $\tilde{M}$ is the estimated number of compartments.
As $\mathrm{P_d}$ does not distinguish between missed fibers and extra compartments found by the reconstruction, we also make use of the following two quantities where needed, $\mathrm{n^-}$ and $\mathrm{n^+}$, explicitly counting the number of \emph{under-} and \emph{over-estimated} compartments, respectively.

%  METRIC 4
\item \emph{Angular error}. This metric quantifies the angular accuracy in the estimation of the directions of the fiber populations in a voxel:
\begin{equation}
	\epsilon_{\theta} = \frac{180}{\pi} \, \arccos( \, | \vect{d} \, \cdot \, \tilde{\vect{d}} \, | ) ,
\end{equation}
where $\vect{d}$ is a true direction and $\tilde{\vect{d}}$ is its closest estimate.
The final value is an average over all fiber compartments \MOD{by first matching the estimated directions to the ground-truth without using twice the same direction}.

\end{itemize}

Peaks detection was performed using a local maxima search algorithm on the recovered FOD, considering a neighborhood of orientations within a cone of $15^{\circ}$ around every direction. 
For this reason, evaluation metrics are not sensitive for small crossing angles and results are reported in a conservative range $30^\circ\textrm{--}90^\circ$.
To filter out spurious peaks, values smaller than $10\%$ of the largest peak were discarded; in the case of \Ltwo we had to increase this threshold to $20\%$, as suggested in \citet{Tournier:2007aa}, in order to compare with the other methods.

%%%%%  Real data  %%%%%
\subsection{Real data} \label{RealData}

The human brain data have been acquired from 3 young healthy volunteers on a 3T Magnetom Trio system (Siemens, Germany) equipped with a 32-channel head coil using standard protocols routinely used in clinical practice. Each dataset corresponds to a distinct subject.
Two DTI scans (referred in the following as \DataDTIa and \DataDTIb) were acquired at $b = 1000 \; \mathrm{s}/\mathrm{mm}^2$ using 30 and 20 diffusion gradient directions, respectively, uniformly distributed on half the unit sphere using electrostatic repulsion \citep{Jones:1999aa}.
Other acquisition parameters were as follows: $\mathrm{TR}/\mathrm{TE} = 7000/82 \; \mathrm{ms}$ and spatial resolution = $2.5 \times 2.5 \times 2.5 \; \mathrm{mm}$ for dataset \DataDTIa, while $\mathrm{TR}/\mathrm{TE} = 6000/99 \; \mathrm{ms}$ and spatial resolution = $2.2 \times 2.2 \times 3 \; \mathrm{mm}$ for dataset \DataDTIb.
One HARDI dataset (referred as \DataHARDIa) was acquired at $b = 3000 \; \mathrm{s}/\mathrm{mm}^2$ using 256 directions uniformly distributed on half the unit sphere \citep{Jones:1999aa}, $\mathrm{TR}/\mathrm{TE} = 7000/108 \; \mathrm{ms}$ and spatial resolution = $2.5 \times 2.5 \times 2.5 \; \mathrm{mm}$.
To study the robustness of the three algorithms to different under-sampling rates, the \DataHARDIa dataset has been retrospectively under-sampled and two additional datasets (\DataHARDIb and \DataHARDIc) have been created, consisting of only 50 and 20 diffusion directions, respectively. These subsets of directions were randomly selected in order to be as much equally distributed on half the unit sphere as possible.
The actual SNR in the $b = 0$ images, computed as the ratio of the mean value in a region-of-interest placed in the white matter and the standard deviation of the noise estimated in the background, was about 60 in \DataDTIa, 30 in \DataDTIb and 30 in \DataHARDIa.

%%%%%  Implementation Details  %%%%%
\subsection{Implementation details} \label{ImplementationDetails}
In all our experiments, the \textit{response function was estimated from the data} following the procedure described in \citet{Tournier:2007aa}.
A different response function was estimated for every combination of experimental conditions (number of samples, b-value, SNR), which was then used consistently in the three reconstruction methods.
Specifically, the 300 voxels with the highest fractional anisotropy were selected as expected to contain only one fiber population, and a tensor was fitted from the dMRI signal in each.
In the case of numerical simulations, an additional set of data containing 300 voxels with a single fiber compartment was generated for this scope.
The estimated coefficients were then averaged to provide a robust estimation of the signal profile for the response function.
As we used the tool \texttt{estimate\_response} of \texttt{mrtrix} for these operations, the estimated kernel was already suitable to be fed into the \Ltwo algorithm.
\MOD{Note that the fiber directions rely on a maxima identification from the SH coefficients, which can take any continuous position on the sphere.}
Conversely, in the case of both \Lone and \Lzero, \MOD{the estimated kernel} was used to create the dictionary $\Phi$ by rotating it along 200 orientations uniformly distributed on half the unit sphere.
\MOD{Because of this discretization, the resulting grid resolution is about $10^\circ$ and thus the intrinsic average error when measuring the angular accuracy is about $5^\circ$.
In other words, the precision of both \Lone and \Lzero is limited by the resolution of the grid used to construct the dictionary. % mentioned in section 2.7
For this reason differences between methods below this threshold will be considered not significant.
Note that, to improve the precision it would be sufficient to increase the number of directions of the discretization which, however, would have serious consequences on the efficiency and stability of the minimization algorithm. Interestingly, recent works of \citet{Tang:2012aa} and \citet{Candes:2012aa} explored a novel theory of CS with continuous dictionaries, in the context of which FOD peaks could be thought to be located with infinite precision. This topic will be the subject of future research.}
Finally, in order to model adequately any partial/full contamination with cerebrospinal fluid (CSF) that may occur in real data, an \textit{additional isotropic compartment} has been considered by adding a column to $\Phi$.
This compartment was estimated by fitting an isotropic tensor in voxels within the lateral ventricles. % which contain only CSF.

\MOD{The} free parameter controlling the degree of regularization had to be estimated for both \Ltwo and \Lone algorithms.
For the former we used the default values suggested in the original implementation available in the \texttt{mrtrix} software.
For the latter, the regularization parameter $\beta$ was empirically estimated following the guidelines of \citet{Landman:2012aa}, in order to place the method in its best conditions. In numerical simulations, we created an additional training dataset for every combination of experimental conditions (number of samples, b-value, SNR) and 50 reconstructions were performed varying the parameter $\beta$ from $10^{-4} \; \beta^*$ to $\beta^*$, with $\beta^* = || 2 \Phi^T \vect{y}||_\infty$ computed independently in each voxel. The value providing the best reconstructions (according to the above metrics) was then used to run \Lone on the actual data used for the final comparison. We did not observe any improvement in the reconstructions outside this range.
\MOD{In real data, we tested different values for $\beta$ but, as the ground-truth is unknown, the optimal value was chosen on the basis of a qualitative inspection of the reconstructions considering their shape, spatial coherence and adherence to anatomy. Nonetheless, we found that the algorithm was quite robust to the choice of $\beta$ and the value providing visually the best results was always very close to $\beta = 0.1 \cdot \beta^*$, as suggested in the same work. Therefore this value was used in all real data experiments. This stability might be probably due to the adaptive strategy of estimating $\beta^*$ in each voxel from the signal $\vect{y}$. }
\MOD{As already emphasized}, \Lzero \textit{does not require any free parameter} to be tuned. In fact, in numerical simulations $k$ can be fixed \MOD{in all iterations} to 3 while we can safely assume $k=5$ in real data, hence larger than the 2-3 fibers normally assumed.

%%%%%%%%%%%%%%%%%%%%%%%%%%%%%%%%%%%%
%%%%%  Results and discussion  %%%%%
%%%%%%%%%%%%%%%%%%%%%%%%%%%%%%%%%%%%
\section{Results and discussion} \label{Results}

% Numerical simulations
% %%%%%%%%%%%%%%%%%%%%%
\subsection{Numerical simulations}
We quantitatively compared the three approaches on synthetic data with the aim of assessing the impact on the reconstructions of each regularization scheme (i.e. $\ell_2$, $\ell_1$ and $\ell_0$ \MOD{priors}) under controlled conditions.
In particular, the quality of the reconstructions was evaluated using the metrics introduced above and selectively varying (i) the number of samples and (ii) the b-value of the acquisition scheme, (iii) the noise level and (iv) the crossing angle between the fiber compartments.
Results are reported independently for each experimental condition.

% Volume fractions and $\ell_1$ norm
% ----------------------------------
\subsubsection{Volume fractions and \texorpdfstring{$\ell_1$ norm}{Lg}} \label{VolumeFractionsAndL1}

\begin{figure}[ht]
\centering
\includegraphics[width=11.5cm]{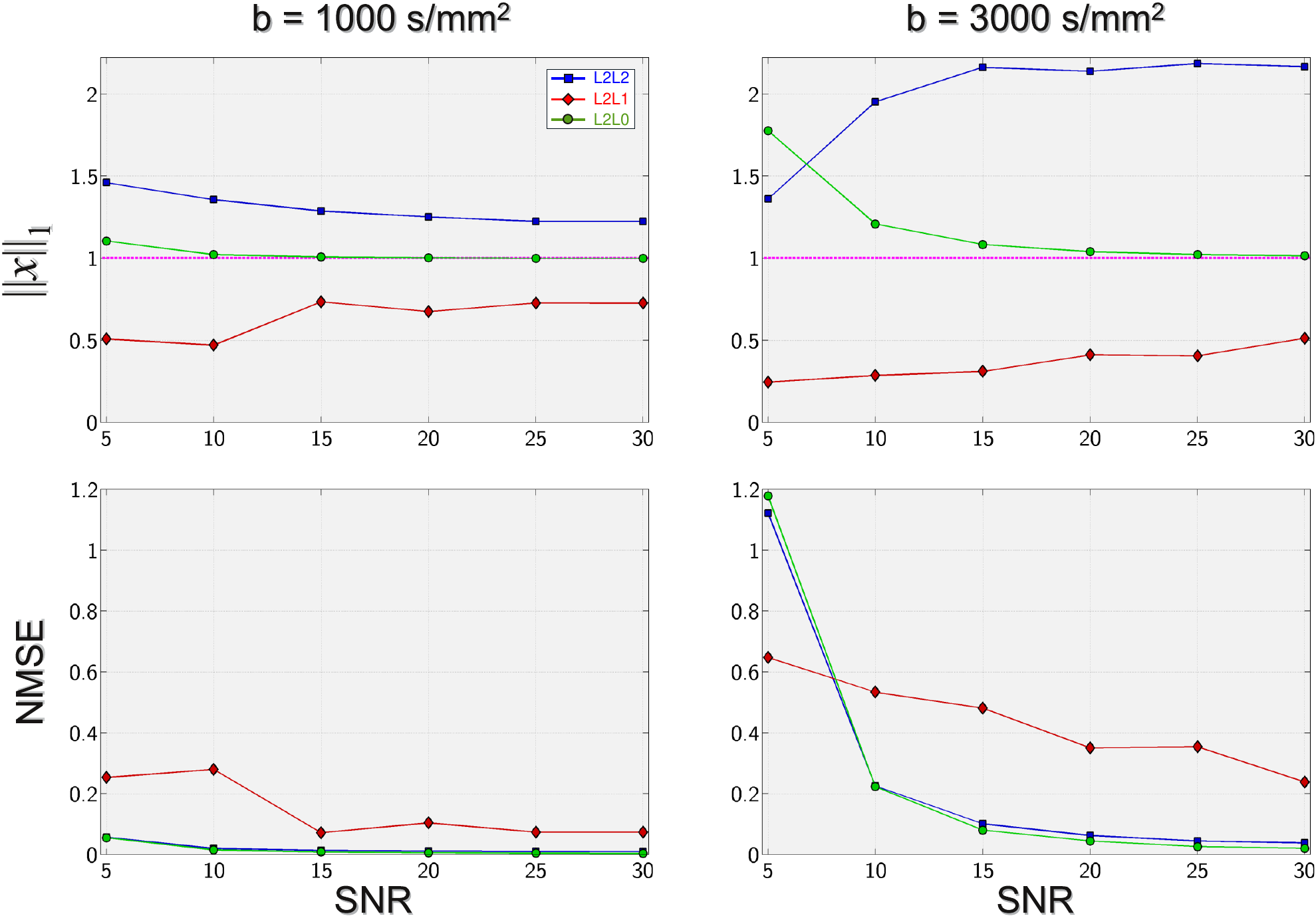}
\caption{\textit{Sum of volume fractions and impact on the reconstructions}. Top plots report the $||\vect{x}||_1$ of the FODs reconstructed by \Ltwo (blue boxes), \Lone (red diamonds) and \Lzero (green circles), while the NMSE of the recovered signal is shown at the bottom.
The reference value $||\vect{x}||_1 = 1$ is plotted in magenta.
Results are reported as a function of the SNR in 2 experimental settings with 30 samples: $b = 1000 \; \mathrm{s/mm^2}$ (left) and $b = 3000 \; \mathrm{s/mm^2}$ (right).}
\label{fig:SumX}
\end{figure}

As previously stated, the physical constraint that the volume fractions sum to unity is normally omitted in every problem formulation, as it is expected to be carried over by the data and properly designed bases \citep{Ramirez-Manzanares:2007aa}.
In Fig.~\ref{fig:SumX} we explicitly tested whether this property is actually satisfied by the algorithms considered in this work.
%The goal here is to draw the attention to an \textit{important side-effect of using different regularization schemes to promote sparsity in the FODs}.
A more detailed analysis of the performance of each prior is performed in the following sections.

The figure reports the average value for the sum of the volume fractions of the reconstructed FODs (i.e. $|| \vect{x} ||_1$), as a function of the noise level, for two acquisition schemes with 30 samples at $b = 1000 \; \mathrm{s/mm^2}$ and $b = 3000 \; \mathrm{s/mm^2}$, respectively.
%The values represent an average over 100 simulations.
The impact on the reconstructions is shown by means of the normalized mean-squared error \MOD{$\mathrm{NMSE = || \vect{y} - \tilde{\vect{y}} ||_2^2 / || \vect{y} ||_2^2}$ \citep{Michailovich:2011aa} between the measured signal $\vect{y}$ and its estimate $\tilde{\vect{y}}$}.
The image clearly demonstrates that both \Ltwo and \Lone reconstructions do not fulfill the $\sum_i \vect{x}_i = 1$ physical constraint, as the sum of the recovered volume fractions always tends to be over-estimated by \Ltwo and under-estimated by \Lone.
\MOD{This is a clear effect of the  weakness of the sparsity constraint in the \Ltwo approach and of the inconsistency of the $\ell_1$ prior in \Lone.}
On the contrary our \Lzero approach appears to correctly satisfy the constraint, with deviations from unity only with very high noise levels (SNR $\approx$ 5).
With high quality data this over/under-estimation behavior is fairly mild (at SNR = 30, $||\vect{x}||_1 \approx 0.7$ for \Lone and $||\vect{x}||_1 \approx 1.2$ for \Ltwo), but it progressively intensifies as the noise level increases.
The trend is even amplified with high b-value data, in which case the $||\vect{x}||_1$ can be as high as $\approx 2.1$ for \Ltwo and as low as $\approx 0.25$ for \Lone.

Despite showing quite different behaviors with respect to the $||\vect{x}||_1$, \Ltwo and \Lzero exhibit very similar NMSE values.
On the contrary, \Lone shows significantly higher reconstruction errors than both \Ltwo and \Lzero, pointing to the aforementioned \textit{$\ell_1$ inconsistency}.
Debiasing methods~\citep{Zou:2006aa} have been proposed with the aim to correct the magnitude of the recovered coefficients and mitigate this effect. Nonetheless, a critical step for applying these techniques consists in the proper identification of the support of the solution, otherwise this procedure can lead to really bad results. As we will show in the next sections, this is the case in this work, as the three methods differ significantly in their ability to estimate the number of fiber populations.
As the very same data and reconstruction basis have been used for all the methods, we can conclude that any deviation from the unit sum has to be attributed to the different regularization employed in each algorithm.
In the following \textit{we will investigate the consequences on the reconstructions} of using different regularization schemes.

% Comparison as a function of the number of samples
% -------------------------------------------------
\subsubsection{Comparison as a function of the number of samples}
\begin{figure}[ht!]
\centering
\includegraphics[width=11.6cm]{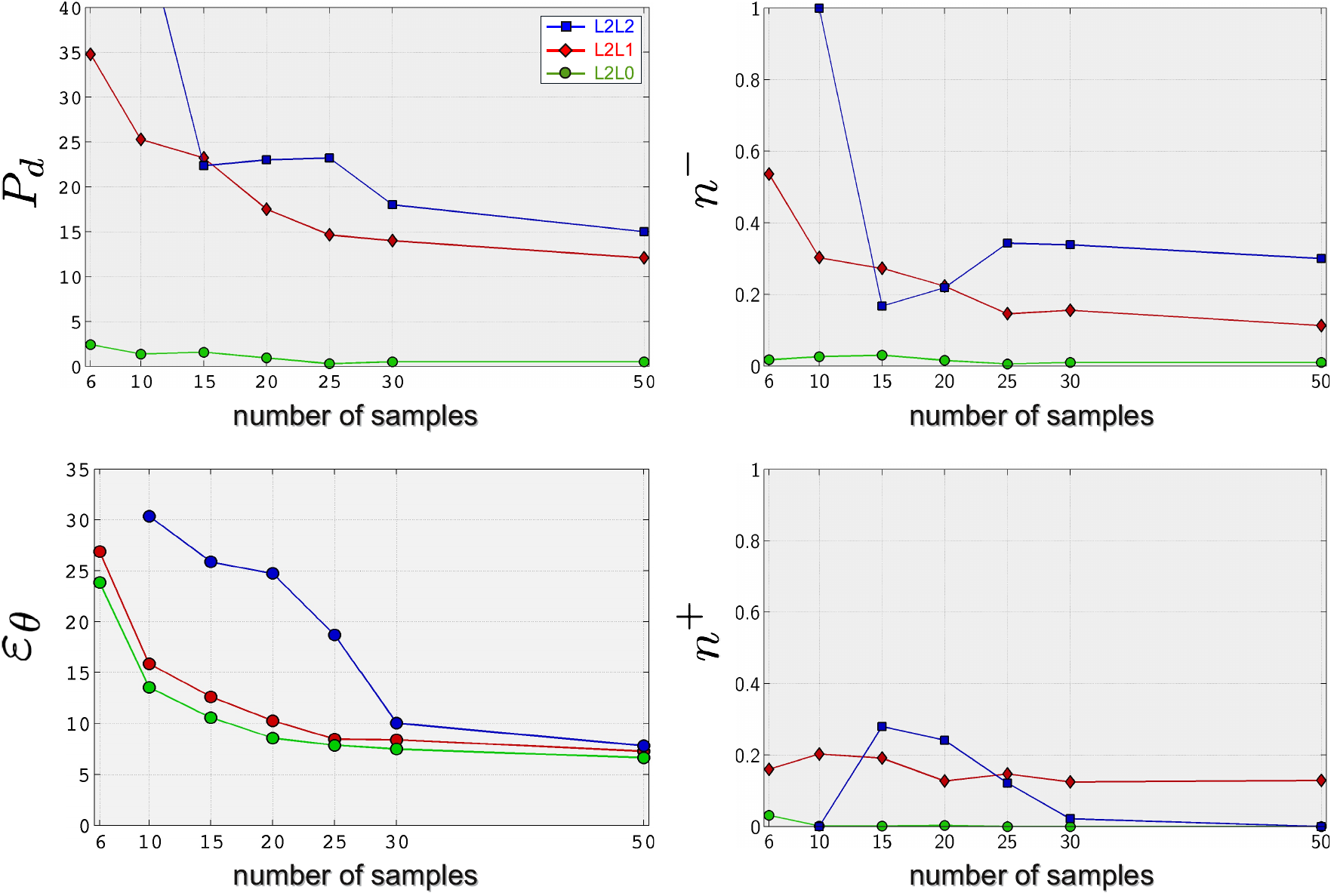}
\caption{\textit{Quantitative comparison as a function of the number of samples.} The values of the four quality metrics are reported for \Ltwo (blue boxes), \Lone (red diamonds) and \Lzero  (green circles) as the number of samples changes. Values shown here correspond to an experimental setting with $b = 2000 \; \mathrm{s/mm^2}$ and $\mathrm{SNR} = 25$.}
\label{fig:performances-vs-nS}
\end{figure}

Fig.~\ref{fig:performances-vs-nS} reports the performance of the three reconstruction methods as the number of samples changes. We considered seven acquisition schemes from 6 to 50 samples and results are reported for a standard scenario, specifically a shell at $b = 2000 \; \mathrm{s/mm^2}$ with a $\mathrm{SNR} = 25$.
The dependence on the b-value and the robustness to noise will be investigated in detail in the following sections.
The quality metrics are reported here as the average value computed over all simulated crossing angles ($30^\circ\textrm{--}90^\circ$).

Looking at the plots the benefits of using an $\ell_0$ prior are clear: \Lzero always outperforms both \Ltwo and \Lone in identifying the correct number of fiber populations ($P_d$)  and results are consistent for all number of samples considered.
The main benefit of \Lzero seems to be the drastically decreased number of missed fibers (smaller $n^-$), even though also the number of over-estimated compartments ($n^+$) is significantly reduced.
Concerning the angular accuracy of the recovered fiber populations ($\epsilon_\theta$), reconstructions with \Lzero always resulted in smaller errors as compared to both \Ltwo and \Lone.
Although the difference with respect to \Lone is \MOD{not significant as always within the intrinsic grid precision}, both methods showed a substantial improvement over \Ltwo, which appeared to suffer from a sudden and significant deterioration of the reconstructions ($\approx 10^\circ\textrm{--}15^\circ$) for less than 30 samples.
This can be explained with the SH representation used internally by \Ltwo. In fact, even though the FOD is a function on the sphere containing high-resolution features by definition, a maximum SH order $l_{max}=4$ (or less) can be used for acquisitions with less than 30 samples, hence drastically reducing the intrinsic angular resolution of the recovered FOD. At least 30 to 60 samples are normally advised for using \Ltwo, so in our experiments we have actually tested \Ltwo beyond its applicability range. On the contrary, \Lone and \Lzero do not make use of SH and the reconstruction quality degrades more smoothly with the under-sampling rate of the dMRI data.
In the following we will focus on two acquisition schemes to further analyze the performance of three methods: (i) in a normal setting with 30 samples and (ii) in a regime of high under-sampling with only 15 samples.

% Comparison as a function of the crossing angle
% ----------------------------------------------
\subsubsection{Comparison as a function of the crossing angle}
\begin{figure}[pt!]
\centering
\includegraphics[width=\textwidth]{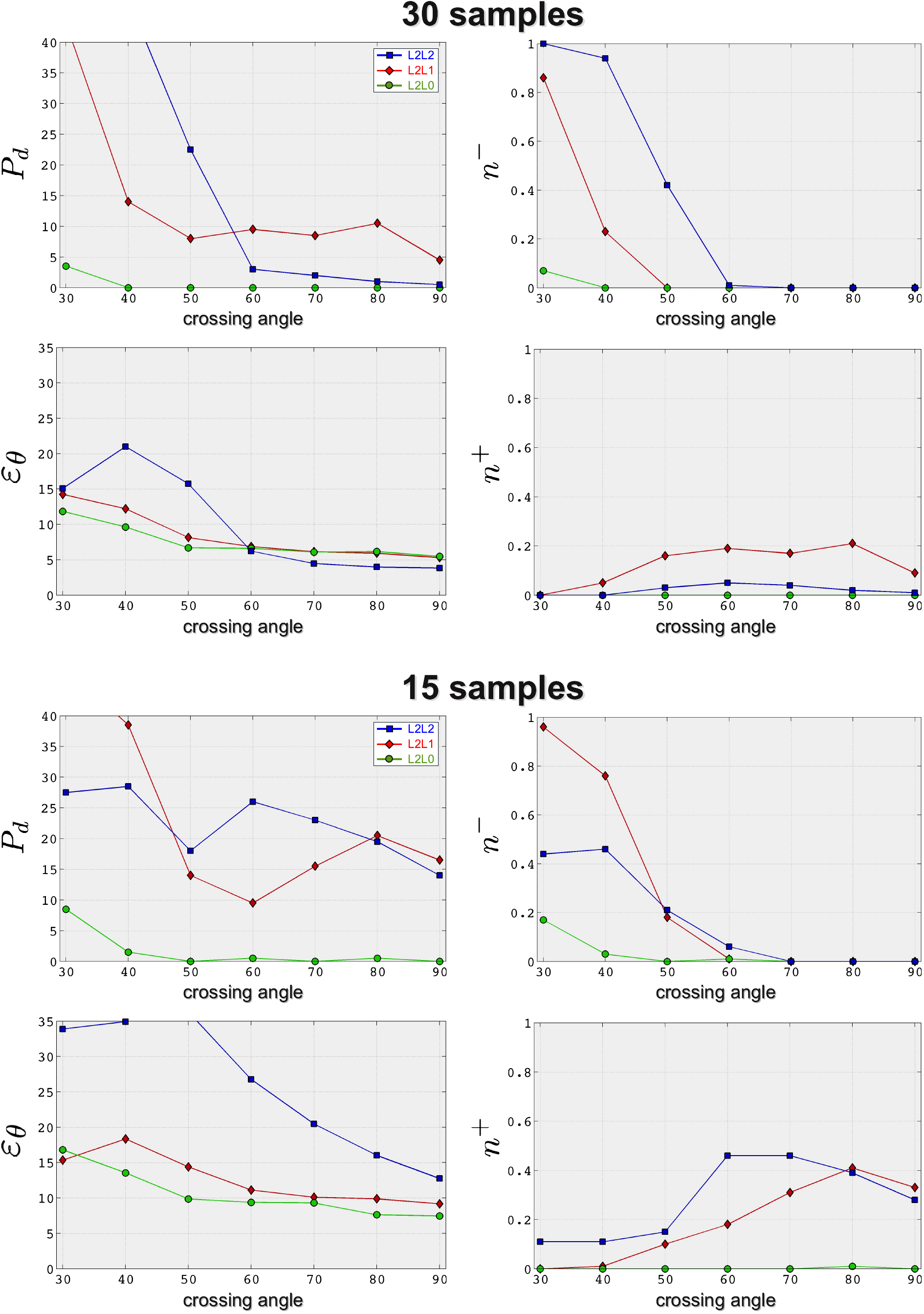}
\caption{\textit{Quantitative comparison as a function of the crossing angle.} The performances of the three reconstruction methods are detailed separately for each crossing angle used in the simulations. Results are reported for 30 and 15 samples, using the same experimental configuration of Fig.~\ref{fig:performances-vs-nS}, i.e. $b = 2000 \; \mathrm{s/mm^2}$ and $\mathrm{SNR} = 25$.}
\label{fig:performances-vs-CA}
\end{figure}

In Fig.~\ref{fig:performances-vs-CA} the performances of \Ltwo, \Lone and \Lzero are plotted in detail as a function of the crossing angle between the fiber populations. 
Results are shown for two acquisitions with 30 and 15 samples, both simulated at $b = 2000 \; \mathrm{s/mm^2}$ and $\mathrm{SNR} = 25$.

With 30 samples, the major source of errors for both \Ltwo and \Lone is represented by \textit{under-estimation} ($n^-$), although spurious orientations are not negligible ($n^+ \approx 0.2$).
In particular, both methods start to severely miss fibers for crossing angles below $60^\circ$, where they tend to recover a single peak lying between the two real fiber directions. In these situations, the maximum angular error for the sole estimated peak is generally upper bounded by half the angle separating the two fibers; for this reason the overall $\epsilon_\theta$ performances of \Ltwo and \Lone do not differ significantly from \Lzero despite the drastic improvement in terms of $P_d$, $n^-$ and $n^+$.
On the other hand, in an under-sampling scenario with 15 samples \Ltwo and \Lone exhibit much higher $P_d$ values and a stronger tendency to \textit{over-estimate} compartments, usually in completely arbitrary orientations not even close to the true fiber directions. The overall improvement in the angular accuracy of \Lzero is more evident, with an average enhancement up to $5^\circ$ with respect to \Lone, whereas \Ltwo exhibits a severe drop of the performance mainly due to modeling limitations, as previously pointed out.

These differences can have dramatic consequences for fiber-tracking applications.
In fact, tractography algorithms are particularly prone to these estimation inaccuracies, i.e. number and orientation of fiber populations, because the propagation of these (perhaps locally small) errors can lead to completely wrong final trajectories. For instance, a missed compartment might stop prematurely a trajectory, while a spurious peak might lead to create an anatomically incorrect fiber tract.
Hence, the ability to accurately recover the intra-voxel fiber geometry is of utmost importance.

% Comparison as a function of the b-value
% ---------------------------------------
\subsubsection{Comparison as a function of the b-value}
\begin{figure}[ht!]
\centering
\includegraphics[width=11.2cm]{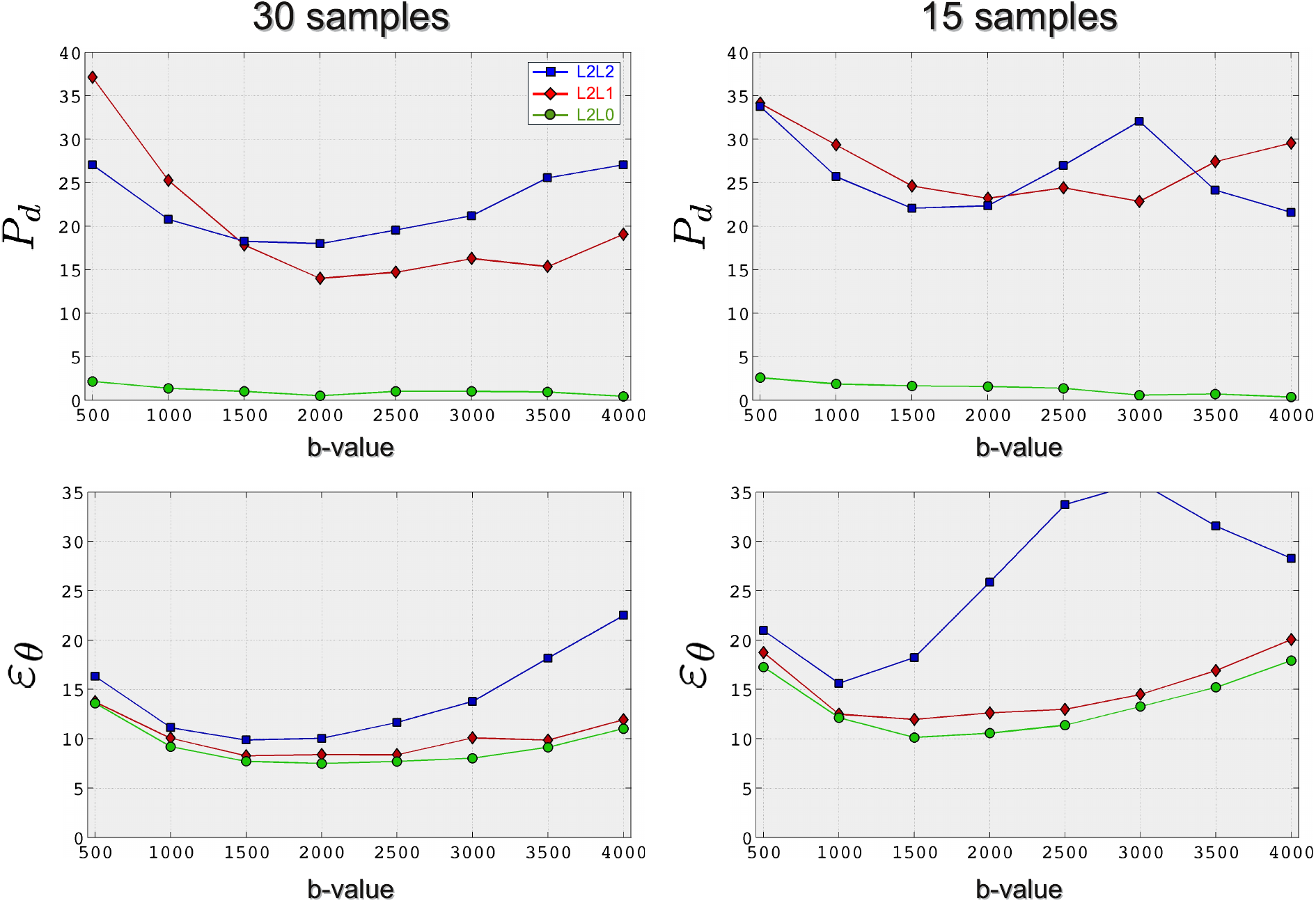}
\caption{\textit{Quantitative comparison as a function of the b-value.} The dependence of the reconstruction quality on the b-values used in the acquisition is reported here for 30 and 15 samples with a $\mathrm{SNR} = 25$.}
\label{fig:performances-vs-b}
\end{figure}

So far \Ltwo, \Lone and \Lzero have been compared for given acquisition schemes at a fixed $b = 2000 \; \mathrm{s/mm^2}$.
Fig.~\ref{fig:performances-vs-b} reports the quality of the reconstructions with the three approaches as a function of the b-value. The results are shown for 30 and 15 samples with a $\mathrm{SNR} = 25$.

\Ltwo tends to miss compartments for low b-values and over-estimate them at higher $b$ ($n^+$ and $n^-$ are not shown here for brevity). 
This is even more apparent when decreasing the number of samples in the acquisition to 15, where \Ltwo estimates a lot of spurious peaks at high b-values (high $n^+$) and thus the angular accuracy of the estimated fiber directions drops considerably.
Interestingly, \Lone shows the opposite behavior, under-estimating at high b and over-estimating at low b, although at a smaller rate thus preventing the performance to degrade significantly.
Again, in comparison, \Lzero shows a very stable estimation of the number of fibers.
Concerning the angular accuracy, all methods showed a minimum for $\epsilon_\theta$ corresponding to $b \approx 1500-2500 \; \mathrm{s/mm^2}$, representing a sort of trade-off between the loss in angular resolution happening at small b-values and the stronger noise influence at higher $b$. In fact, as in this work we report the noise level as the SNR of the $S_0$ dataset, images at high b-values will have lower actual signal-to-noise ratio, and thus the noise effects will be inherently stronger.
Overall, \Lzero always results in smaller angular errors than the other two methods.
The improvement with respect to \Lone is not \MOD{significant}, while the difference with \Ltwo is \MOD{much} more pronounced (up to $20^\circ$) especially as the b-value increases.

% Comparison as a function of the SNR
% -----------------------------------
\subsubsection{Comparison as a function of the SNR}

\begin{figure}[htb]
\centering
\includegraphics[width=\textwidth]{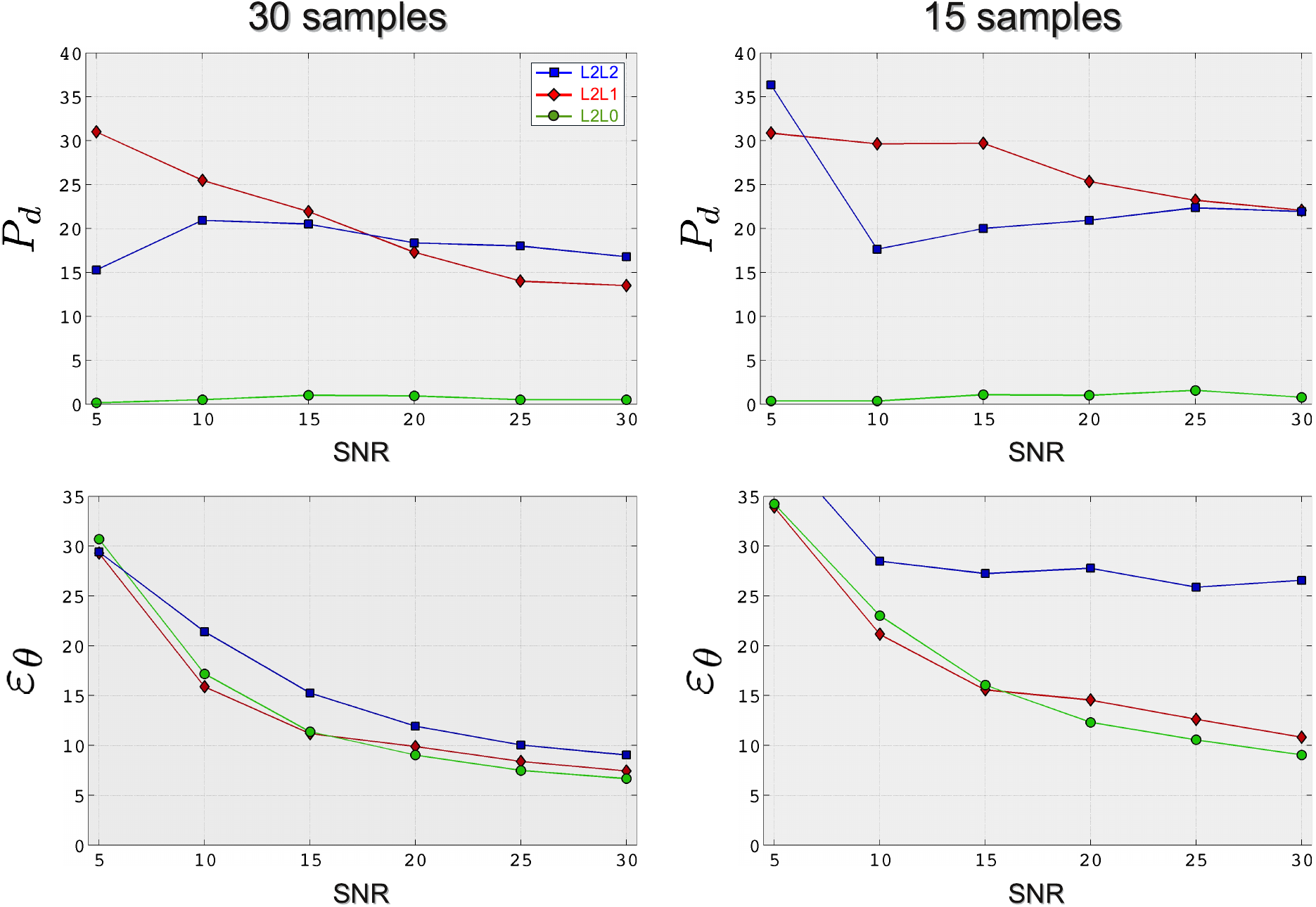}
\caption{\textit{Quantitative comparison as a function of the SNR.} The robustness to noise of the three reconstruction methods in shown for 30 and 15 samples at $b = 2000 \; \mathrm{s/mm^2}$. Reported values for the SNR correspond to the signal-to-noise ratio of the $S_0$ dataset.}
\label{fig:performances-vs-SNR}
\end{figure}

Finally, Fig.~\ref{fig:performances-vs-SNR} compares the robustness to noise of the three methods. Six noise levels have been considered, with the SNR of the $S_0$ dataset varying from 5 to 30. The comparison is reported for 30 and 15 samples at $b = 2000 \; \mathrm{s/mm^2}$.
The results show that \Lzero clearly outclasses the other two methods concerning the estimation of the number of compartments ($P_d$) and results are consistent as the SNR changes, both with 30 and 15 samples.
In terms of angular accuracy, \Lzero and \Lone have very similar $\epsilon_\theta$ performances, almost indistinguishable from one another.
%, with \Lzero performing slightly better for $\mathrm{SNR} \geq 15$ and \Lone for $\mathrm{SNR} \leq 10$.
On the contrary, \Ltwo systematically obtains significantly higher $\epsilon_\theta$ values at all considered SNRs (up to $6^\circ$ with 30 samples).
In a high under-sampling regime (right plots), the angular accuracy drastically degrades in the case of \Ltwo and it appears almost independent of the noise level.
This is again consistent with the limitations of the SH representation for acquisitions with very few samples.

% Real data
% %%%%%%%%%
\subsection{Real data}

\begin{figure}[p!]
\centering
\includegraphics[height=17cm]{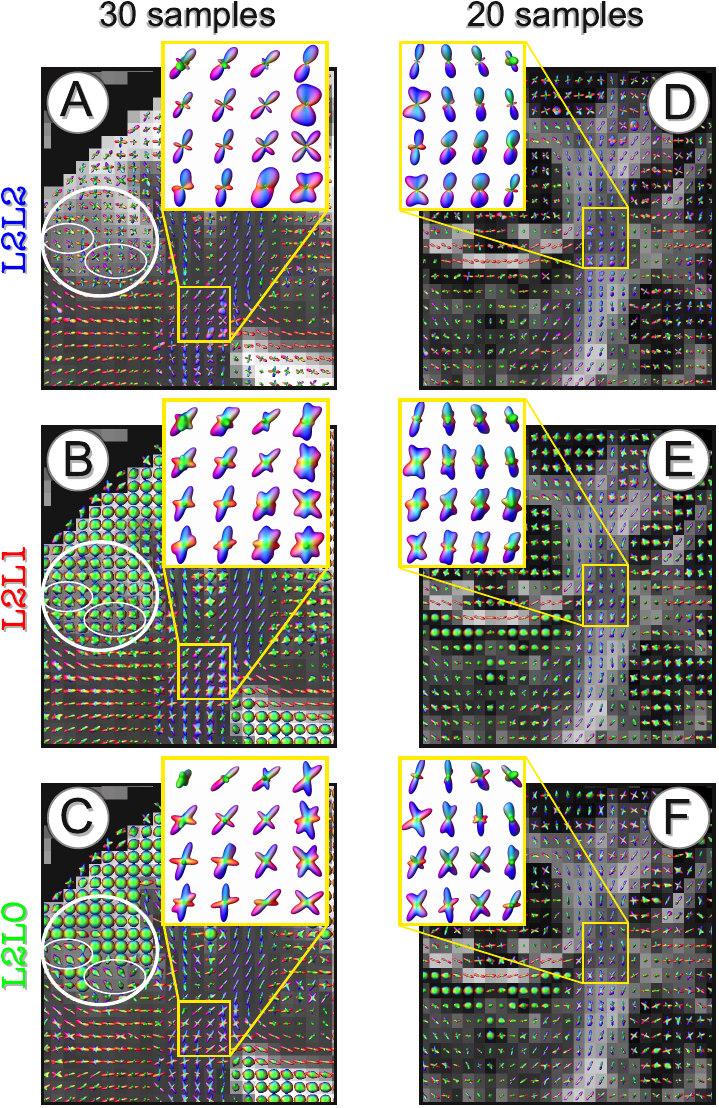}
\caption{\textit{Qualitative comparison on DTI human brain data.} Reconstructions of the FODs in the corona radiata region are shown for: \Ltwo (\textsf{A} and \textsf{D}), \Lone (\textsf{B} and \textsf{E}) and \Lzero (\textsf{C} and \textsf{F}). FODs in subplots \textsf{A--C} correspond to dMRI images acquired using 30 samples, superimposed on the ADC map, while \textsf{D--F} are relative to the acquisition with 20 samples, superimposed on the FA map. All images have been acquired at $b = 1000 \; \mathrm{s/mm^2}$.}
\label{fig:RealData}
\end{figure}

\begin{figure}[p!]
\centering
\includegraphics[width=\textwidth]{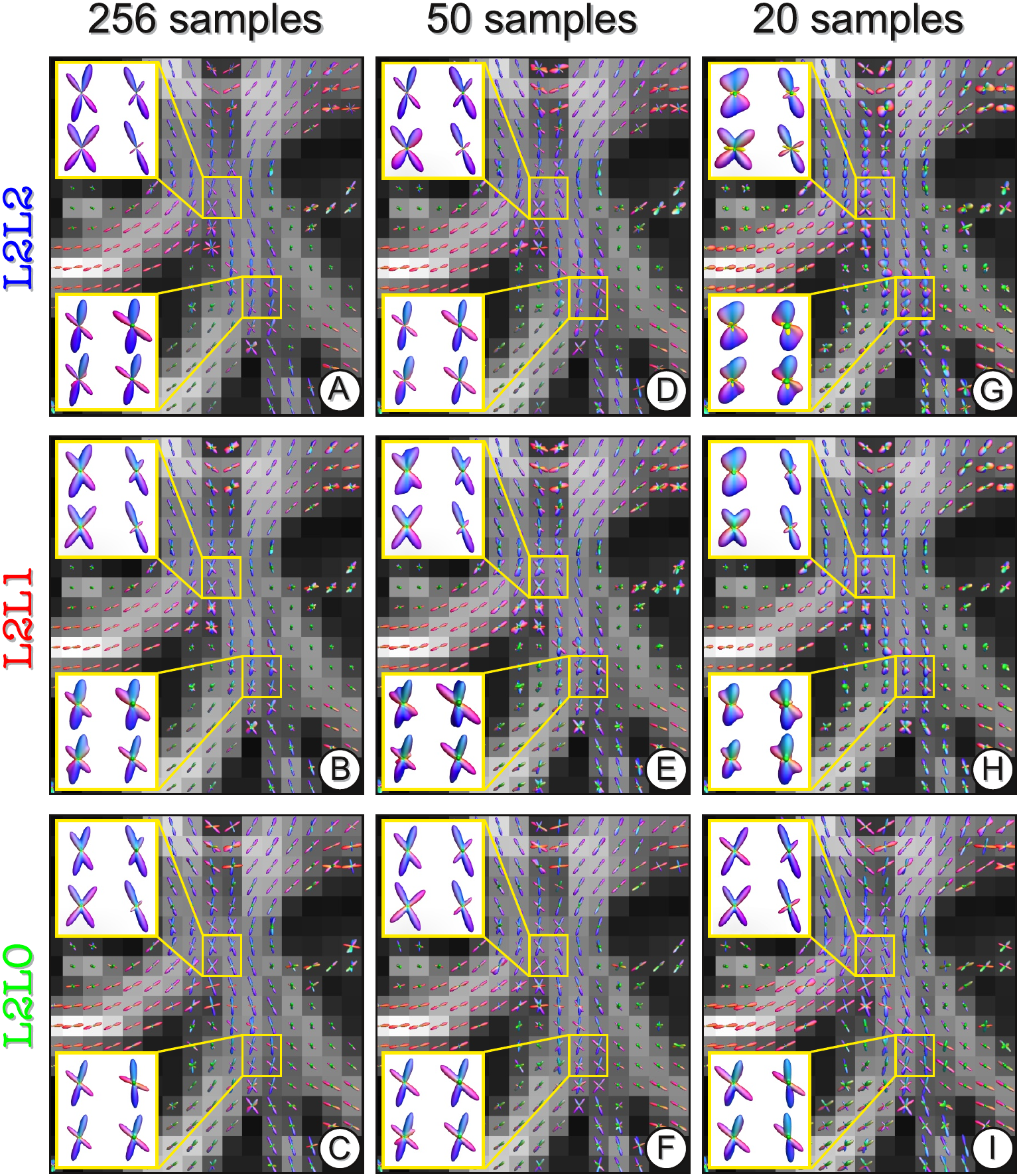}
\caption{\textit{Qualitative comparison on HARDI human brain data.} Reconstructions of the FODs in the corona radiata region are shown for: \Ltwo (\textsf{A}, \textsf{D}  and \textsf{G}), \Lone (\textsf{B}, \textsf{E} and \textsf{H}) and \Lzero (\textsf{C}, \textsf{F} and \textsf{I}). Subplots \textsf{A--C} correspond to the fully-sampled dataset \DataHARDIa (256 samples), \textsf{D--F} to the dataset \DataHARDIb (50 samples) while \textsf{G--I} are relative to \DataHARDIc (20 samples). Images have been acquired at $b = 3000 \; \mathrm{s/mm^2}$.}
\label{fig:RealDataWithHighBValue}
\end{figure}

\subsubsection{\MOD{Qualitative evaluation on DTI data}}

\MOD{Fig.~\ref{fig:RealData} compares the reconstructions\footnote{\MOD{The images have been created using the tool \texttt{mrview} of \texttt{mrtrix}. As a consequence, the FODs from \Lone and \Lzero had to be converted to SH, and this operation caused some blur in the sparse reconstructions of these two methods.}} obtained with the three regularization schemes in the case of real data acquired with a typical DTI protocol.}
Subplots \textsf{A}, \textsf{B} and \textsf{C} correspond to the \DataDTIa dataset acquired using 30 samples.
Even though the acquisition scheme used for this dataset is not the setting where our numerical simulations highlighted the most substantial differences between the three methods, important conclusions can be drawn in favor of \Lzero.
Looking at the regions in the white circles, the ability of both \Lone and \Lzero to properly model the isotropic compartment in voxels with full or partial contamination with CSF is clearly visible.
On the contrary, as \Ltwo does not explicitly model any CSF compartment, it appears unable to adequately characterize the signal in these cases, but it rather approximates any isotropic contribution with a set of random and incoherent fiber compartments.
Besides, comparing \textsf{B} and \textsf{C} we can observe that \Lzero successfully differentiates gray matter (light gray regions) from CSF voxels with pure isotropic and fast diffusion (very bright areas), whereas \Lone appears unable to distinguish them.

The yellow frames highlight the corona radiata, a well-known region in the white matter containing crossing fibers.
As expected from our simulations at this still relatively high number of samples, differences are not obvious between the three methods.
However, we observe that \Lzero clearly results in sharper and more defined profiles than \Lone, whereas the improvements with respect to \Ltwo are confined only to few voxels. The not so good performance of \Lone might be related to the value chosen for $\beta$. In contrast, no free parameter has to be empirically optimized in our approach.
When decreasing the acquisition samples to 20 (subplots \textsf{D}, \textsf{E} and \textsf{F} corresponding to \DataDTIb dataset), \MOD{fiber directions are definitely much better resolved with \Lzero than with both \Ltwo and \Lone}.
In fact \Ltwo clearly breaks, missing many fiber compartments probably due to the aforementioned limitations of the SH representation. The same happens to \Lone, whose reconstructions appear very blurred and noisy.

\subsubsection{\MOD{Qualitative evaluation on HARDI data}}

The comparison with high b-value data is reported in Fig.~\ref{fig:RealDataWithHighBValue}. The figure shows also the robustness to different under-sampling rates of each scheme.
Subplots \textsf{A}, \textsf{B} and \textsf{C} correspond to the fully-sampled dataset \DataHARDIa.
In this situation, no evident differences between the three approaches can be observed as they perform essentially the same.
With moderate under-sampled data (subplots \textsf{D}, \textsf{E} and \textsf{F} corresponding to \DataHARDIb) both \Ltwo and \Lzero do not show any significant difference in the quality of the reconstructions, so far exposing neat and sharp profiles.
On the other hand, the FODs reconstructed by \Lone show some signs of progressive degradation, appearing a little more blurred as compared to those reconstructed from fully-sampled data (compare subplots \textsf{E} and \textsf{B}).
The situation changes drastically with highly under-sampled data, as easily noticeable by comparing the subplots \textsf{G}, \textsf{H} and \textsf{I}, which correspond to the reconstructions performed with only $8\%$ of the original data.
In fact, while \Lzero does not show yet any significant degradation of the FODs, both \Ltwo and \Lone clearly do not provide as sharp and accurate reconstructions as in the case of fully-sampled data (compare \textsf{G} to \textsf{A} and \textsf{H} to \textsf{B}).
In addition, in the case of \Ltwo we can observe a higher incidence of negative peaks (identified in the plots by small yellow spikes), a clear sign of augmented modeling errors.

% Volume fractions and $\ell_1$ norm
% ----------------------------------
\subsubsection{\MOD{Quantitative comparison: volume fractions and \texorpdfstring{$\ell_1$}{Lg} norm}} \label{RealDataVolumeFractionsAndL1}

\begin{figure}[htb]
\centering
\includegraphics[width=10cm]{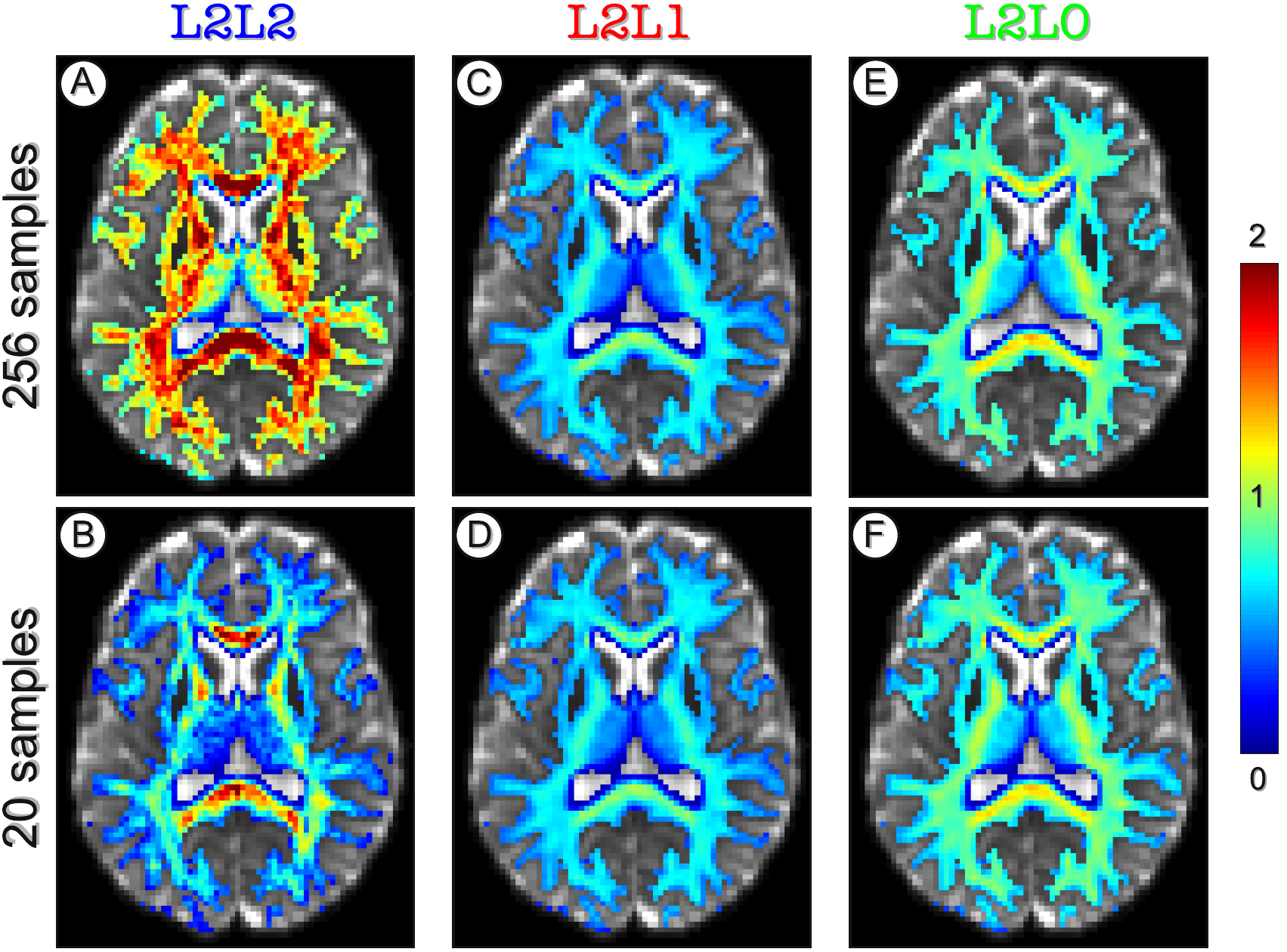}
\caption{\MOD{\textit{Sum of volume fractions in real data}. The sum of the volume fractions of the FODs reconstructed with \Ltwo (\textsf{A} and \textsf{B}), \Lone (\textsf{C} and \textsf{D}) and \Lzero (\textsf{E} and \textsf{F}) is reported in a representative slice of the high b-value data. The top row corresponds to the fully-sampled dataset (\DataHARDIa) while the bottom to the under-sampled one with 20 samples (\DataHARDIc).}}
\label{fig:RealDataSumx}
\end{figure}

\MOD{In Fig.~\ref{fig:RealDataSumx} we tested whether the physical constraint of unit sum is satisfied also in case of real data.
The images confirm the observations previously made with synthetic data (cf. Fig~\ref{fig:SumX}).
In fact, the sum of the recovered volume fractions tends to be over-estimated by \Ltwo (subplots \textsf{A} and \textsf{B}) and under-estimated by \Lone (subplots \textsf{C} and \textsf{D}), whereas \Lzero reconstructions (subplots \textsf{E} and \textsf{F}) appear to meet the property of unit sum as expected.
All methods coherently show a mild over-estimation in the corpus callosum, compatible with the highly-packed axonal structure in this region.
%\Ltwo and \Lone show a marked under-estimation for white-matter voxels close to the gray-matter, especially with a low number of samples (subplots \textsf{B} and \textsf{D}). On the contrary, Lzero seems to model
Finally, \Ltwo seems to suffer from over-estimation more with fully- than with under-sampled data, which might be related to the SH order employed for different number of samples.}

% Quantitative comparison on real data
% ------------------------------------
\subsubsection{\MOD{Quantitative comparison: fully- vs under-sampled data}} \label{RealDataQuantitative}
%number and orientation of fiber compartments

\begin{figure}[htb]
\centering
\includegraphics[width=10cm]{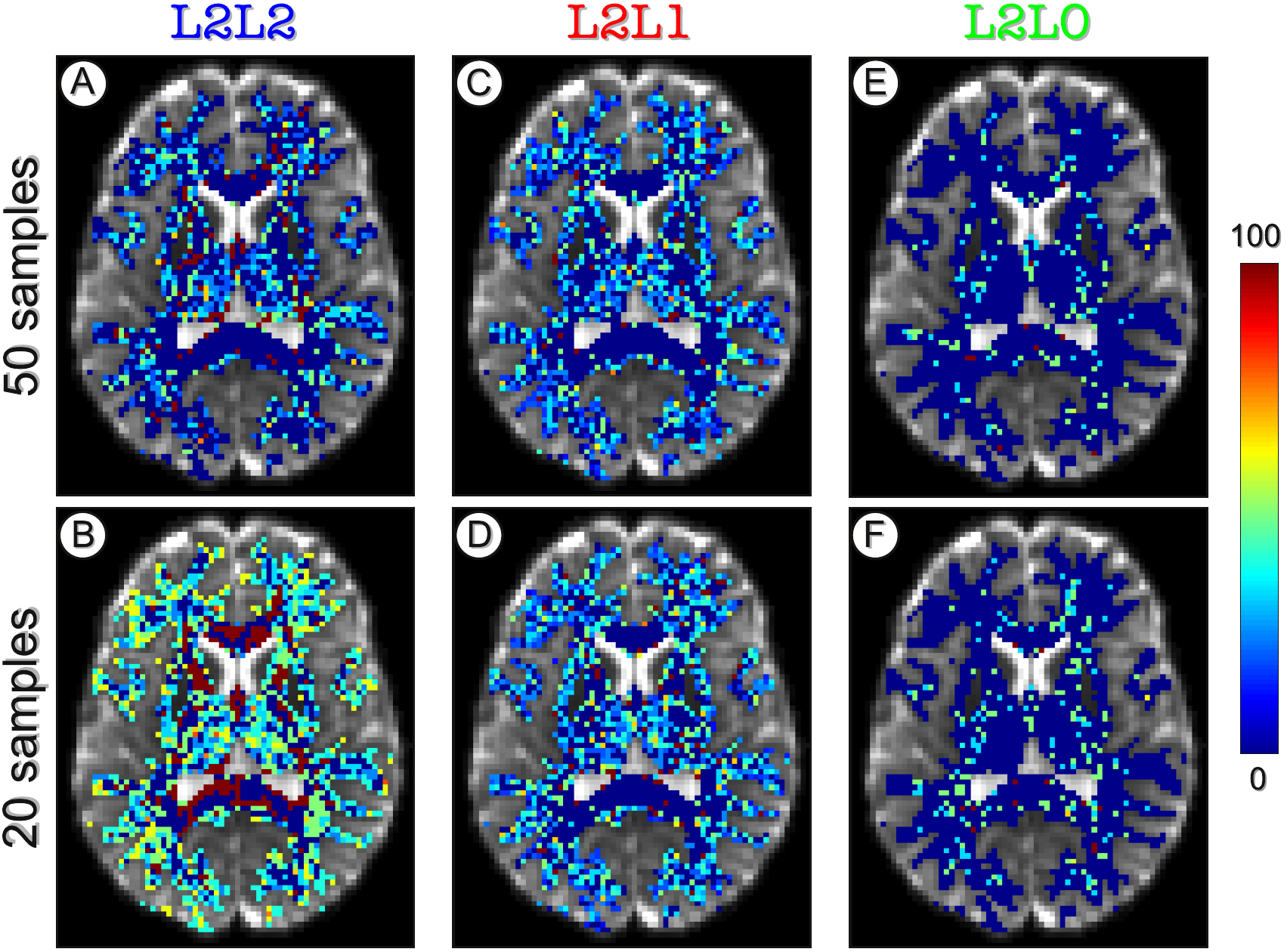}
\caption{\MOD{\textit{Comparison between fully- and under-sampled real data}. The performance of \Ltwo (\textsf{A} and \textsf{B}), \Lone (\textsf{C} and \textsf{D}) and \Lzero (\textsf{E} and \textsf{F}) was quantified by considering \DataHARDIa as ground-truth and computing $P_d$ for the reconstructions on under-sampled data. The top row corresponds to the dataset with 50 samples (\DataHARDIb) and the bottom to 20 (\DataHARDIc).}}
\label{fig:RealDataQuantitative}
\end{figure}

\MOD{We compared the reconstructions obtained from under-sampled data (i.e. \DataHARDIb and \DataHARDIc) to those with fully-sampled data (i.e. \DataHARDIa), considering this latter as ground-truth, as done by \citet{Yeh:2013aa}.
In agreement with the results from numerical simulations, no significant difference was found between the three approaches in terms of angular accuracy.
The average error using 50 samples was $10.9^\circ \pm 9.9^\circ$ (mean $\pm$ standard deviation) for \Ltwo, $8.8^\circ \pm 8.1^\circ$ for \Lone and $10.0^\circ \pm 11.3^\circ$ for \Lzero.
The reconstructions using 20 samples clearly showed higher angular errors. The differences between \Lone and \Lzero were below the resolution of the sphere discretization used in this study: $11.6^\circ \pm 9.1^\circ$ and $12.6^\circ \pm 12.4^\circ$ respectively.
\Ltwo revealed significantly higher $\epsilon_\theta$ values: $17.2^\circ \pm 12.8^\circ$.}
\MOD{On the other hand, results definitely confirmed the superior performance of \Lzero in terms of $P_d$ that was previously observed in synthetic experiments.
With 50 samples \Lzero had an average $P_d = 4.0\% \pm 13.9\%$ as opposed to sensibly higher values for \Ltwo and \Lone, respectively $17.8\% \pm 32.6\%$ and $17.3\% \pm 24.3\%$.
For 20 samples, the performance of \Ltwo and \Lone visibly deteriorated, $42.1\% \pm 43.6\%$ for the former and $21.3\% \pm 27.7\%$ for the latter. \Lzero reconstructions appeared very stable with an average $P_d = 5.5\% \pm 15.8\%$. These enlightening results are illustrated in Fig.~\ref{fig:RealDataQuantitative}.}

%\begin{table}[!h]
%\centering
%\footnotesize
%\setlength\extrarowheight{3pt}
%\begin{tabular}{l|c|c|c||c|c|c|}
%	\multicolumn{1}{c}{} & \multicolumn{3}{c}{50 samples} & \multicolumn{3}{c}{20 samples} \\
%	\cline{2-7}
%	\multicolumn{1}{c|}{} & \textcolor{Blue}{\Ltwo} & \textcolor{Red}{\Lone} & \textcolor{Green}{\Lzero} & \textcolor{Blue}{\Ltwo} & \textcolor{Red}{\Lone} & \textcolor{Green}{\Lzero} \\
%	\cline{2-7}
%	$P_d$ & $17.8 \pm 32.6$ & $17.3 \pm 24.3$ & $\mathbf{\phantom{0}4.0 \pm 13.9}$ & $42.1 \pm 43.6$ & $21.3 \pm 27.7$ & $\mathbf{\phantom{0}5.5 \pm 15.8}$ \\
%	\cline{2-7}
%	$\epsilon_\theta$ & $10.9 \pm \phantom{0}9.9$ & $\phantom{0}8.8 \pm \phantom{0}8.1$ & $10.0 \pm 11.3$ & $17.2 \pm 12.8$ & $11.6 \pm \phantom{0}9.1$ & $12.6 \pm 12.4$ \\
%	\cline{2-7}
%\end{tabular}
%\centering
%\caption{\MOD{Comparison between fully- and under-sampled real data.}}
%\label{tab:RealDataQuantitative}
%\end{table}

% Limitations and future work
% %%%%%%%%%%%%%%%%%%%%%%%%%%%
\subsection{Limitations and future work}

Our proposed \MOD{formulation} represents an extension of classical spherical deconvolution and sparse reconstruction methods \citep{Tournier:2007aa,Landman:2012aa} and, as such, it also inherits all the intrinsic limitations and shortcomings of this class of techniques. 
Like all its predecessors, in fact, our method is based on the assumption that the response function can be estimated from the data and especially that it is adequate for characterizing the diffusion process in all the voxels of the brain.
Moreover, the validity of these approaches has yet to be properly assessed with more critical intra-voxel configurations \citep{Sotiropoulos:2012aa} or pathological brain conditions.
Yet, as these methods are currently widely used in this field, we have shown in this work that by \textit{expressing adequately the regularization prior used for promoting sparsity} the quality of the reconstructions can significantly be improved, with no additional cost.

Some of the aforementioned limitations might be addressed by enhancing the estimation of the dictionary accounting for more complex configurations, such as using different response functions for different brain regions and/or pathological tissues and including specific kernels which explicitly model fiber fanning/bending.
In addition, even though we focused here on single voxel experiments, future work will be devoted to study the applicability and the effectiveness of our approach in more sophisticated frameworks exploiting the spatial coherence of the data.
Finally, future research will investigate the use of the recently proposed continuous CS theory \citep{Tang:2012aa, Candes:2012aa} with the aim of further improving the accuracy of the reconstructions and reducing the acquisition time.

%%%%%%%%%%%%%%%%%%%%%%%%
%%%%%  Conclusion  %%%%%
%%%%%%%%%%%%%%%%%%%%%%%% 
\section{Conclusion}\label{Conclusions}

In this paper we focused on \MOD{spherical deconvolution} methods currently used in diffusion MRI for recovering the FOD and estimating the intra-voxel configuration in white matter.
In particular, we investigated the effectiveness of state-of-the-art regularization schemes based on $\ell_2$ and $\ell_1$ priors and provided evidence that these formulations are intrinsically suboptimal: \MOD{the former because it does not explicitly promote sparsity in the FOD, the latter because it is inconsistent with the fact that the fiber compartments must sum up to unity}.
\MOD{We proposed a formulation that rather places a strict bound on the number of expected fibers in the voxel through a bound on the $\ell_0$ norm of the FOD, relying on a reweighted $\ell_1$ scheme}.
We compared our \MOD{\Lzero} approach with \MOD{the} state-of-the-art \MOD{\Ltwo and \Lone methods}, both on synthetic and real human brain data.
Results showed that our proposed formulation significantly improves \MOD{single-voxel FOD} reconstructions, with no additional overheads.
This evolution is most remarkable in a high q-space under-sampling regime, thus driving the acquisition cost of HARDI closer to DTI.

\section*{Acknowledgments}
This work was supported by the Center for Biomedical Imaging (CIBM) of the Geneva and Lausanne Universities, EPFL, the Leenaards and Louis-Jeantet foundations, the EPFL-Merck Serono Alliance award and the Swiss National Science Foundation (SNSF) under grant PP00P2-123438.

%%%%%%%%%%%%%%%%%%%%%%%%
%%%%%  References  %%%%%
%%%%%%%%%%%%%%%%%%%%%%%%
\section*{References}
\bibliographystyle{model2-names}
\bibliography{references}

\end{document}